\documentclass[a4paper,11pt]{article}
\usepackage{setspace}
\usepackage[titletoc,title]{appendix}
\usepackage{amsfonts} % for the \checkmark command
\begin{document}
\title{Blended Learning Content Generation:\\
A Guide for Busy Academics}
\author{Richard Hill\\
	School of Computing and Engineering,
	University of Huddersfield\\
	\texttt{r.hill@hud.ac.uk}\\
	Version 1.1}
\date{June 2020}
\maketitle
\tableofcontents
%
%\begin{abstract}
%This article describes ...
%\end{abstract}
%
\section{Introduction}
This guide is an attempt to summarise the concepts that we should consider when producing teaching resources for online or blended learning delivery. It is intended to support the transfer of traditional face-to-face (F2F) resources, into an engaging learning experience, where access to on-campus facilities is either severely limited or completely absent.

%Any removal of F2F content means that we need to review what teaching is possible, and it is likely that there may be some new content to generate that embraces online delivery.

Creating new content, for a new delivery method, within a tightly constrained environment can be daunting. But there are ways forward that not only minimise the effort of transferring teaching into new circumstances, but they also maximise opportunities to enhance the learning experience.

\subsection{Terminology}
We should be aware of the relevant terminology when thinking about different modes of delivery. First, we shall consider the following:
\begin{itemize}
    \item{\emph{Synchronous activity}} - interactions that happen in real-time, at the same time. Examples include virtual meetings, telephone calls, real-time chat, onsite teaching/demonstrations in a classroom or lecture theatre, etc.
    \item{\emph{Asynchronous activity}} - interactions that do not occur at the same time. Email conversations are an example, where each interaction is not necessarily responded to within the same time frame. Wikis, discussion forums and written correspondence are other examples of asynchronous activity.
\end{itemize}
Social media tools or online collaboration platforms can blur the distinction between synchronous and asynchronous interactions; Google Docs permit both real-time interaction (synchronous) and reflective, asynchronous interaction depending upon how they are used.
\subsection{Online or blended?}
Traditional correspondence-based distance learning (DL) courses are examples of \emph{asynchronous} interactions where learners study remotely, in a predominantly self-structured way, leading towards an assessment endpoint. Such courses lend themselves to online delivery; written materials are converted into electronic form and provided via the internet.

The proliferation of online asynchronous courses might give an impression that this is the extent of online learning. However, digital tools that support real-time interaction means that online courses can also include synchronous activity. Allen and Seaman\cite{allen2016} define an online course as having at least 80\% of its content being delivered using digital devices and networks.

Typically, a university course combines on campus delivery of teaching through lectures, seminars and workshops, supported by web-based technologies such as Virtual Learning Environments (VLE), to provide supporting content and services for assessment. Predominantly, the teaching is experienced on site rather than remotely.

A hybrid model, where between 30-79\% of learning activity is delivered online, is referred to as \emph{blended} learning\cite{allen2016}. A key characteristic of blended learning is that the necessity to be on site is reduced from that of traditional teaching approaches.

There is no reduction in the requirement to provide synchronous learning activities in blended courses, and there should be \emph{no reduction} in the number of learning hours for each learner. In many cases, academics who employ blended approaches to teaching have found ways which increase the value of the interactions with and between learners, leading towards improved learner engagement, attainment and satisfaction. 
\subsection{Overview}
Section \ref{sec:keyPrinciples} gives an overview of the pedagogical design implications of more flexible learning materials. All situations requiring new resources, whether existing learning materials are available or not, should consider these content creation and delivery principles.

Section \ref{sec:flexibleLearningVignettes} describes different experiences of modifying learning content and delivery patterns to make the experience more flexible.

Section \ref{sec:summingUp} brings together the key points that should help expedite the adoption of a more flexible approach to university teaching.

A checklist for flexible content generation is included in Appendix \ref{sec:flexibleDeliveryChecklist}.

This guide is deliberately short, so that the key points can be put into practice quickly. It is not meant to be an exhaustive treatment of online or blended learning creation; links to other sources of assistance are included for the reader to engage with in Appendix \ref{sec:furtherReading} if required.

\section{Eight principles for content delivery}
\label{sec:keyPrinciples}
The prospect of designing and delivering an effective online course that is to be delivered wholly or partially online can seem to be a significant challenge. All of the established practices of live, real-time interaction between tutors and students may be reduced or absent, particularly if learners in a cohort reside in different timezones, or if physical facilities limit the volume of learners that can gather in one place.

With online materials we can no longer rely upon a verbal explanation to accompany static, visual presentations. We cannot gauge the understanding of a class and dynamically re-purpose our arguments to foster understanding.

However, most online content developers report that their reconstruction and reflection upon the learning activities for online purposes not only produces a different, but remarkably effective learning experience, but there is the added benefit that their approach to face-to-face (F2F) delivery improves also.
\subsection{Principle 1: Generate learning hours}
\label{subsec:generateLearningHours}
When designing F2F modules, there is a tendency to quantify the learning experience in terms of delivery or `contact' hours. For a single term this might be a 1 hour lecture and a 2 hour tutorial per week, for 12 weeks. This equates to 36 hours, out of a budget of 200 hours for a 20 CAT module. The remaining 164 hours are typically attributed to `guided independent study'.

\begin{quote}When thinking about online or blended delivery, it is useful to consider the amount of \emph{learning hours} generated rather than the number of hours of \emph{contact}.
\end{quote}

An online learner may be attempting to fit their study around other commitments, and therefore would appreciate an indication of the amount of notional time they need to allocate to a learning activity.

We should not overload learners; there is a risk that we can unwittingly prompt them to engage in activities, that when considered as a whole, amount to considerably more effort than we intended.

Online, asynchronous courses can generate huge workloads for learners if the individual tasks are not thought about at the outset. We also know that for a given piece of content (say a set of presentation slides), the rate at which the learner digests the learning may be variable.

It is useful to prompt ourselves with questions such as:
\begin{itemize}
	\item How long would it take to answer a question about ...?
	\item Can we realistically expect the learner to complete that task in X hours?
\end{itemize}
In a traditional lecture, the person presenting the slides governs the delivery rate. Depending upon the complexity of material, the detail on each slide, the number of questions posed, and a variety of other factors, the pace of delivery is managed to fit the overall time allocation. This cannot be controlled in the same way in an asynchronous situation, and it is debatable whether recorded lectures successfully support learning that effectively anyway.

It is far better to consider the questions that need to be posed, to reinforce the learning outcome that is desired. This is best achieved by thinking about the learning hours that our materials generate, rather than worrying about replicating X number of \emph{contact} hours.

Once we consider learning hours, we can start to think about learning activities, posing questions such as:
\begin{itemize}
    \item How can we prompt learners to find solutions to problems?
    \item Can we facilitate their learning rather than delivering content to them?
    \item How can we help learners locate content that is of use to them?
\end{itemize}
In fact, after thinking about how people really learn, you may decide not to use long-form lecture presentations at all.
\subsection{Principle 2: Provide structure}
%
% need to talk about course goals and the need to relate exercises to goals - starting with the end in mind, etc.
%
One of the most noticeable aspects of asynchronous courses for learners is the apparent lack of structure. The materials may be fastidiously organised into a sequence and presented just as the academic intended the content to be studied, but an asynchronous course lacks the discipline and schedule of lectures, seminars, workshops and tutorials.

As such, the introduction of asynchronous content can result in the removal of structure, or at least a reduction in the real-time touch-points where learners can interact to learn from staff and each other.

We can help address this issue by explicitly stating what the aims and outcomes of the module are, and how we are going to use the materials during the course of the learning.

\begin{quote}
Official module documents are not always an integral part of flexible teaching resources, so make sure that the learning outcomes are properly contextualised in the introductory text that you provide for the learners.
\end{quote}

We must consider this when designing our materials. We know that there are greater numbers of learners who must balance their studies alongside paid employment, and also caring responsibilities.

So, if we have planned our content from the perspective of generating learning hours (Principle 1), then why should we not let our learners know this?
\begin{quote}
    \textit{To prepare for the next online discussion, you should investigate the current state of the art developments in formal verification methods for Cyber Physical Systems (CPS).}
    
    \textit{\textbf{Task 1:} Using the university's online research repository, find an academic journal article that describes an example of model verification syntax which has been applied to a CPS.}
    \textit{You should use the SQ3R method to record the details of your search.}

    \textit{This task should take \textbf{1 hour to complete}}.
\end{quote}
Providing learners with an indicative amount of study time helps them structure the learning activities around their other commitments. It also makes it clear what we expect of them.

If they struggle to complete it within the guided time allocation, that can be a topic to be discussed during a small group or personal tutoring session.

Those weekly prompts to engage, via scheduled classes that are delivered onsite, synchronously, also need to be replaced with another activity that provides a similar, structured prompt to engage.

It might be that instead of interacting on site, a video conferencing tool can be used to bring together the learners into a synchronous, virtual learning space. This may or may not be suitable for your teaching interactions. Such an interaction might not be suitable when the group grows beyond a certain number of participants.

In such cases we need to re-think what interactions need to take place to maximise the learning opportunities. For instance, while we have been teaching classes in-person in a computer lab for some years now, is this still the best way of teaching what needs to be learned?\\

Which interactions need to be F2F \emph{and} on campus?\\

What if your classroom was removed or its seating capacity was reduced?\\

How would you overcome such circumstances?\\

Reflecting on what we need to deliver can help us think about the structure we need to provide for our learners, and how we can align effective learning experiences to that structure\cite{biggs2007}.

If we can deliver a better experience by combining two or three asynchronous activities with a synchronous activity (that may be online or on-campus), then we can simultaneously improve the delivery of the learning, while also enhancing our learners' opportunities to be successful.
\subsection{Principle 3: Design to the lowest common denominator}
Even in these modern times of reasonable-to-high network bandwidth and high-specification personal computers, smartphones and tablets, it is prudent to design content that can be consumed with meagre resources.

Video should \textit{only} be utilised when it is crucial for the learning experience. Learners report that they often prefer audio recordings such as podcasts, which can be consumed more easily within busy lives, such as when commuting, for instance. Audio content is relatively easy to produce, and most likely takes less time to prepare than lecture slides.

Learners don't like timed presentations with slides that automatically advance; they prefer to navigate at their own pace, pausing, reviewing and repeating the material to aid their understanding. Similarly, they find animations frustrating unless the animation serves to reinforce understanding.

So, think about the `action' that can be generated by simple text, augmented by diagrams/drawings/images, and some prompt questions as appropriate, to stimulate the learner to engage and learn.
\subsection{Principle 4: Create reusable learning `chunks'}
Modularity is an important topic for digital content developers. When we develop content, it should provide some substantial time-savings when we re-use that content somewhere else. For example, we could use it as a supplementary piece of learning for a F2F, on-campus learner to complete as part of their `guided independent learning'.

Complications tend to arise when F2F materials are `converted' to online materials (or \textit{learning objects}). Often, the result is a set of materials that align with the F2F delivery (12 weeks of content for instance), with some prompt questions thrown-in for good measure. The problem now is that unless another module requires a week's worth of content, an individual learning object is closely coupled to the original module. As a consequence the learning object cannot be re-used, and any potential for efficiency gain is lost.

Efficiency through re-use is very important for learning materials, since a large investment in time is required initially. One way to ensure that flexible learning is expensive is to create new content for every instantiation.

We avoid this by firstly concentrating upon \textit{learning hours} (Principle 1), using this as a guide to control the number of activities that will be required. We can then concentrate upon creating learning objects that convey discrete pieces of learning, which are usually smaller than the typical on-campus lecture or tutorial, within which 2, 3 or more concepts may be conveyed during a single session.

Thus, we should approach the design of online materials from the perspective of separate learning objects that can be assembled into a learning experience of a certain length, rather than mapping topics to week/block/ term/semester F2F delivery.
\subsection{Principle 5: Assess frequently and with purpose}
In a teaching session a tutor asks many questions to stimulate and reinforce understanding. Within the session learners ask and also answer their own questions. When creating learning content we often forget the volume and intensity of questioning that takes place in the classroom.

One way of addressing this could be to use multiple choice quizzes (MCQ), which are often a part of e-learning and VLE applications. These can provide quick formative feedback (a good thing). If we invest some additional time, we can provide model-answer explanations as to why some options are incorrect.

However, such quizzes should not be viewed as a universal solution. What about open-ended questions? These can set the scene for some learner-centred activity, particularly if you task the learners to report back via a discussion forum, or comment on a blog posting, or even a reflective summary that they submit as part of a logbook or portfolio.

Similarly, time-constrained activities don't have to be limited to MCQ or tests; learners can be tasked to solve a problem either individually or as a group, to provide an answer within a time limit.

If we find ourselves struggling to find time to use a specialist facility (or specific equipment), then we need to think about how we can support the learners to a) experience deliberate practice using the specialist resources, and b) show us only what is absolutely essential to be demonstrated with the equipment, and thus, assess other aspects using alternative methods.

For example, a particular learning outcome requires some `hands-on' guided practice with specialist equipment. Questions we might pose about the design of suitable learning activities are:
\begin{itemize}
    \item Does a group of learners need to be observed using the facility or can learners work unsupervised?
    \item How can structured, unsupervised, deliberate practice be supported and enriched by formative assessments?
    \item What specific aspects of the skill require physical contact? How can we assess those skills within the time constraints?
    \item What portions of the practice can be replaced with simulations or virtual environments?
    \item To what degree are we assessing outcomes over process?
\end{itemize}
Thinking about constraints, and how these might change the assessment practices that have existed before, can be an effective way of driving creativity. If we generate learning hours, how can our time be best utilised to support and develop authentic assessments? 
\subsection{Principle 6: Help learners teach each other}
There is the irony that it is the lecturer, who has to explain concepts, and re-frame those explanations to suit different learners, who learns the most when teaching. But, traditional lectures concentrated on the broadcasting of content to passive audiences, for learners to use as a reference point for further study. 

Fortunately, a lot of lectures have moved away from this and become much more interactive, as academics understand that learners can perform better with this approach.

Online scenarios are interesting as it is often assumed there will be asynchronous communication through the use of a discussion forum. Learners can exchange messages and engage in discussions about topics, which are grouped as threads.

As tutors we can exploit this environment to foster much deeper, more effective learning, by teaching the learners to teach their peers. This reinforces their learning of the prescribed topics, enabling them to bring their own learning and experiences into the module.

These teaching interactions produce \textit{unintended} learning outcomes that are over and above the planned experience. We can help learners by posing questions for them to respond to, but also by asking them to critique each others' work.

We might provide initial prompt questions to get started, facilitating deeper discussion by offering targeted, strategic input as a thread of discussion develops.

Think about exercises in formative assessment that can be `triggered' by poignant questions - how can you develop your learning material to take full advantage of learners teaching each other?
\subsection{Principle 7: Assess process \emph{and} product}
There is an argument that remote learning is difficult to assess, since there is the challenge of ensuring the authenticity of the `virtual' learner. How do we know who is actually sitting the assessment? This is a problematic area for many academics.

We can consider assessment \textit{for} learning as opposed to the more traditional assessment \textit{of} learning. Whilst this is a relatively subtle change of words, it has much more impact upon the learning experience.

Assessment for learning prompts the learning content developer to create more opportunities for assessment to take place, and to enable the learner to derive more benefit earlier in the teaching delivery. It promotes the use of feedback (from either tutors, peer learners or preferably both) to direct and refine a learner's contributions, to enhance their outputs and demonstrate first-hand the processes of enquiry.

VLE quizzes are a rapid way of checking understanding and providing formative and summative feedback on learner enquiry. But, use these with moderation; excessive, repetitive testing can disengage learners. It is wise to think about the variety of ways that feedback can be provided.

For example, setting a task where learners must review their peers' work and provide written comments can demonstrate the real value of feedback. Tutors learn when they provide feedback. Why shouldn't students experience this as well?

Research is an interesting vehicle for promoting learning; we expect researchers to collaborate as they conduct their enquiry. The result is a product of the interactions that each party has engaged with along the way. Learners who master research processes will have learned something extremely valuable, which extends far beyond the information that is `crammed' for an examination, or `researched' for a summative piece of coursework.

If we design online materials where the assessment process (formative and summative) is an integral part (such as a multi-faceted portfolio of artefacts, logbooks, technical reports that have been joint-authored, etc.), then there is a greater likelihood of learner engagement, and a reduced risk of blatant plagiarism (reflections are more difficult to plagiarise consistently and convincingly). The main point is that the iterative assessment approach rewards continued engagement, which is a useful trait of a structured course.

As academics this is an opportunity to bring more of our research behaviours into teaching. Engaged learners can conduct experiments, write research articles and even edit books when they are given the chance. In our stewardship roles, we can create the conditions for learners to work alongside us and develop distinctive, graduate attributes along the way.
\subsection{Principle 8: Evaluate}
Reflection can be a challenging topic to teach. Some learners find it difficult to get to grips with, and struggle to express themselves through writing.

However, asynchronous learning has a much greater emphasis upon the written word, it being the primary communication method. Try and exploit this from the outset, by building in short answer, formative quizzes and discussion forum tasks, into your learning materials.

%More extended reflection can be promoted by having learners discuss solutions to a problem in a discussion forum, or by writing in a personal logbook or blog. Once a series of these tasks have been completed, you can then prompt learners to evaluated the progress that they have made, before the end of the module. This will promote deeper thinking and subsequent understanding.

Something else to consider is `thinking' time. That is, allocating time where the learner is directed to step away from the materials and reflect upon the progress that they are making.

A few simple questions along the lines of ``what have you done, can you apply your new knowledge, what do you now need to know" can easily generate 30 minutes or so of focused learning activity. Make sure that this is factored into your consideration of learning hours (Section \ref{subsec:generateLearningHours}).
\subsection{I'm not sure about my on-campus F2F materials now!}
If we apply these principles to all of our content generation, then there are potential benefits beyond providing engaging online material design. Modularised content can be used to supplement F2F materials.

On campus, your prompt questions could be used in a lecture, treating the occasion as a large interactive session. Campus-based F2F learners often respond well to materials that were originally intended for online learners.

%As staff we should not neglect our own self-evaluation.
Tools that facilitate asynchronous learning such as quizzes, can also be used to solicit feedback on perceptions and experiences of learners. This is helpful for our own self-evaluation.

Just don't expect the learners to complete a weekly questionnaire to ``rate my tutor". This approach is likely to dis-engage learners. Formative feedback is most effective when it is central to the learning activity.

Think about how you can design learning activities that provide the feedback to support evaluation. You can then use this to adapt your delivery as the course progresses.
%Perhaps we have been missing something for some time.

%
\subsection{I can't see where I can make the time to do this}
A primary concern for academic staff is how to find the time to adapt their teaching to become more flexible, so that it can accommodate varying proportions of remote, on-campus, synchronous and asynchronous delivery.

Any course, whether it be fully online or blended, on or off-campus, must stimulate the same amount of work from the learners. There \emph{should not be more effort} required to complete a blended course. We should keep our task allocation in check so that we do not create excessive workloads for learners.

There is also the concern about the process of transforming the teaching resources, to make them more flexible.
If existing materials exist, there might be a temptation to produce a recorded audio soundtrack for each set of lecture sides, and then to open up online discussion sessions for tutorial or workshop sessions.

This is a bad idea on two counts. First, the time taken to record audio onto lecture presentations is likely to take at least 1.5 or even 2 times the duration of the lectures. For a course with 26 hours of lectures, that is an additional 52 hours of speaking, recording and editing.

Second, learners do not respond well to lengthy online presentations. Such materials actively disengage learners and increase the probability that they do not complete their studies.

However, if we think about learning hours rather than contact hours, we can start to explore more time-efficient ways of developing flexible content that actively engages learners.

The existence of lecture slides, or lecture capture recordings opens up the possibility of re-purposing those as resources to be consulted. Task-based activities can be developed that signpost learners to content where required. This can take the form of a set of textual instructions.

Directing a learner to investigate, discuss, solve a problem or provide feedback to a peer, are all examples of instructions that can stimulate considerable learning hours, without incurring a large amount of material preparation hours for time-pressed academic staff. The vignettes in Section \ref{sec:flexibleLearningVignettes} give some examples that avoid the lengthy process of narrating lecture slides.

It is best to reflect upon the curricula and to consider which of the learning interactions are best facilitated by F2F interaction (either on or off-campus) and which of the learning could be delivered by other means.

Student learning occurs via student-student interaction as well as student-tutor interaction; which aspects of your module can explore both of these modes?

\section{Flexible learning vignettes}
\label{sec:flexibleLearningVignettes}
Having examples of the outcomes that we want learners to achieve can be an effective way of demonstrating what is possible. Deciding to change the method of teaching delivery is a significant undertaking and therefore it is useful (and reassuring) to learn from others' experiences.
This section contains a series of vignettes, each describing how a different approach to teaching delivery can transform the outcomes of both learners and academics.

The following example module delivery structures are offered purely as an example of how some of the principles in this guide have been interpreted in the past.

These examples are not intended to be \emph{the answer} for every teaching situation. But, they might inspire some thoughts about how you can re-engineer teaching to suit your learners better.
\subsection{Programming Students Write a Book}
A conversation amongst some academics led to the realisation that while the top-performing final year dissertations won prizes, there were many more excellent pieces of work that went unrecognised each year. As an experiment, one of the academics decided to change their final year module so that it resembled a standard academic publishing project.

The students had one semester of 12 weeks to create and edit a research book. The class met on campus four times throughout the module, and worked almost entirely online using a discussion forum within the VLE to record each stage of their work. 

Students worked in pairs for the first two weeks, at which point they produced an online research poster. The whole cohort provided each other with feedback, after which pairs were asked to join to make small groups of four. Each group was then required to deliver two chapters that related to the ideas generated by the research posters.

The tutor made time available each week to provide feedback to the groups on their work within the discussion forum, as well as delivering weekly tasks to help those that were stuck, or were having difficulties editing. Each week, groups were assigned a different group so that they could provide feedback on the work to date.

At the end of the module, the edited manuscript was assembled and published online.

In twelve weeks of teaching, a total of eight (contact) hours were used for on-site F2F teaching. The tutor spent 12 hours providing feedback on the discussion forum.

The students, who were programmers, had a reputation for finding writing challenging. In twelve weeks they learned about the process of publication, how to use writing prompts, how to write to deadlines, and how to collaborate to produce an extended document of 18 chapters for publication.
\subsection{Zero Curriculum}
Some second-year undergraduate students were given a set of learning outcomes, a selection of VLE tools, and a semester to design their own solution to an industrial problem.

The students elected to meet bi-weekly on campus, and to meet virtually otherwise. They requested staff to be present in weeks 6, 9 and 12. The remainder of the time the staff provided online support through the VLE.

At the end of the module the students delivered their proposal to the industrial client, who was impressed by their use of technology to manage distributed learning over the semester, eventually leading to a solution that achieved tangible savings for the company.

This module required a relatively small amount of preparation, and around 4 hours of on-campus F2F contact, and 12 hours online asynchronous support.
\subsection{Teaching Programming Skills via Job Interviews}
Postgraduate students from the Indian Subcontinent were attracted to a course that required the fast acquisition of software development skills. The module had a poor reputation for not being able to instil the requisite skills in a relatively condensed period of time.

The module was literally turned on its head by examining the motivations of the students; they wanted to secure employment as software developers first and foremost, and the existing module was letting them down.

As a result, the lectures were replaced by a series of interviews, role-played by two members of academic staff. Each week, one academic would ask the other a typical technical question from a job interview for a programming job.

The academic 'interviewee' gave answers that would demonstrate some of the pitfalls of poor preparation, and this enabled a short summary at the end of each week's session that described what the interviewee would need to know. In the first instance, the interviews were role-played live in a lecture theatre, but all sessions were video recorded for online use in subsequent years.

These weekly events, and the associated lists of what the interviewee needed to know, became the template for weekly two-hour workshops. Students were guided to study and practice the topics raised in each of the mini-interviews, and a summative online test was held at weeks 4, 8 and 12 of the one semester course.

The key objective of the transformation was to address the previously dismal 40\% first time pass rate. This immediately became 78\%, and after another year of refinement became 90\%. Perhaps more importantly, the students achieved greater attainment on subsequent modules as a result of being more skilled.
\subsection{Using Formative Quizzes to Transform Attainment}
First year undergraduate students needed to learn introductory statistics and data analysis skills as part of their course. There was a wide range of numeracy abilities present, and the subject was renowned for being challenging to teach, especially maintaining interest from learners.
The module had been delivered over 12 weeks, with one hour of lecture and one hour of tutorial per week.

Academic staff decided to retain the lecture materials and lecture capture videos from previous years, but to use these as materials to refer to, rather than to replay content.

Each week, two key concepts were extracted from the lectures, and a short video was recorded (less than five minutes for each concept) of the academic explaining the concept. A list of two or three tasks, together with the indicative time required to complete them, was then written for each of the concept videos.

Finally, a formative assessment quiz was created in the VLE for each week, which drew relevant questions at random from a question pool.

Learners were encouraged to complete the materials and the quiz each week. In the following week, the tutor ran a scheduled online meeting in the time slot for the lecture, where the results of the formative test for the previous week was discussed during the first 15 minutes of the session.

The key thrust was that the results were presented as statistics, and the cohort discussed how they might analyse the results with the tutor. The remainder of the session was spent discussing questions generated by the concept videos for that week.

Weekly tutorial sessions remained on campus, and were focused upon the application of their knowledge to a case study that ran alongside the weekly virtual discussion.

This change in delivery had a considerable effect upon student engagement and attainment. It became much clearer that students understood the application of statistical ideas much sooner in the course, since they applied and practiced the ideas on their own results.
\section{Summing-Up}
\label{sec:summingUp}
The general advice here is to consider the production of materials for flexible learning as a fundamentally different opportunity to engage learners in learning. Attempts to replicate F2F, classroom-based learning as online or blended delivery, are often quite frustrating for learners.

Voiceovers for MS Powerpoint slides can be useful to explain a concept, but they should not be the basis of the delivery. Similarly, 20-odd sessions of one hour lecture recordings does not make for a great learning experience.

Some time spent thinking about what you want your learners to achieve can result in some remarkably simple, but effective, content that is flexible to consume, and also straightforward to adapt for the future.

A key part of facilitating remote and asynchronous learning is establishing and maintaining learner activity. This can be achieved by specific prompt questions that are nestled amongst the learning content. These questions should prompt learners to reflect, apply, share and critique their learning; you might create a space for such activity in a discussion forum for that topic, for example.

This keeps the discussions in the correct `containers' for each topic, making it easier for tutors to facilitate, and learners to navigate prior experiences.

It is also prudent to populate an empty discussion forum with some simple questions to get the online conversation flowing. These can relate to the questions in the relevant learning objects for each topic.

In summary, when facilitating online and blended learning with digital learning materials, keep Principles 1 (generate learning hours), 6 (help learners teach each other) and 8 (evaluate) at the forefront of your mind.

Appendix \ref{sec:flexibleDeliveryChecklist} has a checklist of items to think about when creating flexible delivery materials This list, alongside this guide, should be seen as a starting point to ensure that the fundamentals are in place.  There are many variations of excellent practice which deviate or build upon the principles described here and Appendix \ref{sec:furtherReading} contains a short list of further reading that is an excellent next step.

%
%\bibliography{mybib}{}

\begin{thebibliography}{}
%
\bibitem{allen2016}
Allen, E., Seaman, J. (2016). \emph{Grade level: Tracking online education in the United States}. The Babson Survey Research Group. Retrieved from: http://onlinelearningsurvey.com/reports/onlinereportcard.pdf
%
\bibitem{biggs2007}
Biggs, J., Tang, C. (2007). \emph{Teaching for quality learning at university}. Maidenhead, UK: Open University Press.
%
\bibitem{bloom1956}
Bloom, B.S. (Ed.). (1956). \emph{Taxonomy of educational objectives handbook I: Cognitive domain}. New York, NY: Longman.


\end{thebibliography}
%\bibliographystyle{101fc2}

%
\pagebreak
\begin{appendices}
\appendix
%\addappheadtotoc
\section{Flexible Delivery Checklist}
\label{sec:flexibleDeliveryChecklist}
Use this checklist as an aide-m\'emoire when designing learning materials for flexible teaching delivery.
\subsection{Course organisation}
\begin{itemize}
    \item[\checkmark] A clear statement of the learning outcomes within the introductory text.
    \item[\checkmark] Learning activities that explicitly link to learning outcomes.
    \item[\checkmark] Activities are clearly identified as being on or off-campus.
    \item[\checkmark] Learning activities are in a logical order/hierarchy that makes it clear what has to be studied when.
    \item[\checkmark] All assessment activities are clearly identified and accompanied by details of how to complete.
\end{itemize}
\subsection{Learning object design}
\begin{itemize}
    \item[\checkmark] Learning materials are `chunked' into appropriately-sized pieces of learning.
    \item[\checkmark] There is a clear thread of activities that are interwoven between synchronous and asynchronous learning materials.
    \item[\checkmark] Learning activities build upon preceding activities to incrementally build difficulty.
    \item[\checkmark] The most appropriate technology is used for each activity: video only where necessary; audio content; text based content.
    \item[\checkmark] Static, text-based content is written in a conversational tone to improve learner engagement.
\end{itemize}
\subsection{Activity management}
\begin{itemize}
    \item[\checkmark] Activities are varied and require the learner to engage with the VLE.
    \item[\checkmark] Learners are prompted to pause and evaluate their own, and their peers' work.
    \item[\checkmark] All learning activities have an indicative workload, specified as the time to spend completing the task.
    \item[\checkmark] Formative and summative assessment tasks are made clear to all learners.
    \item[\checkmark] Learners are required to use the same rubrics as those they shall be assessed by.
    \item[\checkmark] Communication channels are an integral part of the way in which the learning is consumed.
    \item[\checkmark] Each learning outcome has at least one opportunity for formative assessment for the learners to gauge progress.
    \item[\checkmark] Supplementary, non-essential resources (such as further reading) should be marked as such.
\end{itemize}
%

%\pagebreak
%
\section{Further Reading}
\label{sec:furtherReading}
Here is a small selection of resources that can help the interested reader.
\begin{itemize}
    \item \emph{The changing landscape of assessment: some possible replacements for unseen, time-constrained, face-to-face invigilated exams.}
    Professor Kay Sambell, Edinburgh Napier University and Professor Sally Brown, Independent consultant.
https://sally-brown.net/download/3148/
    \item \emph{Essentials for blended learning: A standards-based guide.} Routledge.
    Jared Stein and Charles R. Graham.
    \item \emph{Assessment digest.} Extracts from books by Phil Race.
    https://phil-race.co.uk/download/5589/
\end{itemize}

\end{appendices}
\end{document}